# Transformation of spin information into large electrical signals via carbon nanotubes

Luis E. Hueso[1†], José M. Pruneda[2,3*], Valeria Ferrari[4], Gavin Burnell[1‡], José P. Valdés-Herrera[1,5], Benjamin D. Simons[4], Peter B. Littlewood[4], Emilio Artacho[2], Albert Fert[6] & Neil D. Mathur[1]

[1] *Department of Materials Science, University of Cambridge, Pembroke Street, Cambridge CB2 3QZ, UK*

[2] *Department of Earth Sciences, University of Cambridge, Downing Street, Cambridge CB2 3EQ, UK*

[3] *Institut de Ciencia de Materials de Barcelona, CSIC Campus U.A.B., 08193 Bellaterra, Barcelona, Spain*

[4] *Cavendish Laboratory, University of Cambridge, JJ Thomson Avenue, Cambridge CB3 0HE, UK*

[5] *Nanoscience Centre, University of Cambridge, JJ Thomson Avenue, Cambridge CB3 0FF, UK*

[6] *Unité Mixte de Physique CNRS-Thales ,TRT, 91767 Palaiseau and Université Paris-Sud, 91405 Orsay, France*

[†] *Present address: ISMN-CNR, via Gobetti 101, 40129 Bologna, Italy*
[*] *Present address: Department of Physics, University of California, Berkeley, CA 94720, USA*
[‡] *Present address: School of Physics and Astronomy, University of Leeds, Leeds, LS2 9JT, UK*

**Spin electronics (spintronics) exploits the magnetic nature of the electron, and is commercially exploited in the spin valves of disc-drive read heads. There is currently widespread interest in using industrially relevant semiconductors in new types of spintronic devices based on the manipulation of spins injected into a semiconducting channel between a spin-polarized source and drain[1,2]. However, the transformation of spin information into large electrical signals is limited by spin relaxation such that the magnetoresistive signals are below[2] 1%. We overcome this long standing problem in spintronics by demonstrating large magnetoresistance effects of 61% at 5 K in devices where the non-magnetic channel is a multiwall carbon nanotube that spans a 1.5 μm gap between epitaxial electrodes of the highly spin polarized[3,4] manganite $La_{0.7}Sr_{0.3}MnO_3$. This improvement arises because the spin lifetime in nanotubes is long due the small spin-orbit coupling of carbon, because the high nanotube Fermi velocity permits the carrier dwell time to not significantly exceed this spin lifetime, because the manganite remains highly spin polarized up to the manganite-nanotube interface, and because the interfacial barrier is of an appropriate height. We support these latter statements regarding the interface using density functional theory calculations. The success of our experiments with such chemically and geometrically different materials should inspire adventure in materials selection for some future spintronics.**

We show how carbon nanotubes (CNTs) can be used to solve a long standing challenge in spintronics, i.e. the injection of spins into a non-magnetic material and the subsequent transformation of the spin information into a large electrical signal. This challenge began in 1990 when Datta and Das[5] introduced the concept of the spin transistor. The device is based on spin injection into a semiconductor channel between ferromagnetic contacts and spin manipulation by a gate voltage. In all spin transistor concepts based on similar structures[1,2,6], the common prerequisite for device operation is the existence of a significant magnetoresistance (MR=$\Delta R/R_P$) of the order of unity or larger, where $\Delta R=R_{AP}-R_P$ is the resistance change that results when a magnetic field is used to alter the relative orientation of the magnetizations of the source and the drain between antiparallel (AP) and parallel (P). A recent review[2] of experimental results notes that MR values have been limited to ~0.1-1%, far short of the minimum acceptable ~100%. We demonstrate a 61% effect with a significant output voltage change that can reach 65 mV, and explain why this improvement was made possible by selecting a multiwall CNT to separate source and drain.

CNTs are relatively robust and easy to manipulate, and have been successfully employed[7] in proof-of-principle field effect transistors, quantum dots and logic gates. For spintronics, the weak spin-orbit coupling presents the intrinsic advantage of a long spin lifetime. As explained later, the other key advantage of CNTs is their large[8] Fermi velocity $v_F$, related to the zero band-gap character of the electronic structure and the resulting linear dispersion[7]. However, it is far from obvious whether spin information can survive long-distance transport given the likelihood of defects and contamination.

Our study of CNTs with ferromagnetic electrodes represents a fusion of molecular[9] and spin electronics[1], i.e. molecular spintronics. In this nascent field, MR effects are typically confined to low temperatures in devices based on octanethiol[10], $C_{60}$ [11] or CNTs[12-15]. These CNT devices were electroded with cobalt[12,15], Pd-Ni [13] or GaMnAs [14], and their MR effects were studied at low bias and low temperature. The MR is generally small (~10%) and exhibits inversions of sign, either from sample to sample, or as a function of voltage[12-15]. This is due to complex effects related to Coulomb blockade and level quantization. To remove these effects, we performed our MR experiments up to temperatures as high as 120 K, and at bias voltages larger than 25 mV (which is possible here given that the current is limited by a naturally occurring tunnel barrier at each LSMO-CNT interface). This voltage is sufficient given that the Coulomb blockade energy[8] for similar CNTs is ~0.1 meV, and given also that the level spacing $hv_F/2L$ is ~0.8 meV for an undoped metallic tube of length $L=2$ μm with $v_F = 0.8 \times 10^6 m/s$ [8].

It is not *a priori* known whether spin information can be efficiently transmitted between two materials that possess very different geometries and chemistries. In this Letter, we present devices (Fig. 1 and Methods) where epitaxial electrodes of the pseudo-cubic perovskite manganite $La_{0.7}Sr_{0.3}MnO_3$ (LSMO) are electrically connected by a single multiwall CNT, that lies on top of the electrodes in contrast to standard nanotube device geometries[7]. At low temperatures, the conduction in LSMO exhibits a very high spin polarization[3,4] approaching 100%, whereas elemental ferromagnets commonly used in spintronics possess low spin polarisations[16] of <40%. Moreover, since LSMO is an oxide, it displays environmental stability so that one may attempt to introduce molecules ex-situ.

Similar and reproducible zero-field current-voltage (*I-V*) characteristics (Fig. 2) were seen in 12 devices. Four of these show the large MR effects discussed later, and the other 8 show no MR effects. Our CNT-LSMO interfaces behave like tunnel junctions given that the $I(V)$ curves are strongly non-linear, and given also that the low bias (25 mV), low temperature (5 K) resistance $V/I$ = 10-100 MΩ of our 12 devices is 3-4 orders of magnitude larger than the inverse of the quantum conductance $e^2/h$, i.e. ~13 kΩ typically seen for nanotubes between standard metallic electrodes[13,15,17]. Note that tunnel barriers are generally found at the interfaces between LSMO and metals[18]. However, our interfacial resistance[19] is not unduly high given that we estimate it to correspond to a value falling within the wide range of values[18] associated with metal-LSMO contacts.

The existence of the observed tunnel barriers may be understood in terms of first-principles calculations (Methods) of the electronic structure of a LSMO-CNT interface. The electronic structure of the CNT is not significantly altered when contacted by LSMO (Fig. 3a) suggesting that the barrier is localised at the interface. The Kohn-Sham potential[20] – the simplest estimate of the local energy of a tunnelling electron – shows a barrier (Fig. 3, inset) whose height somewhat exceeds the characteristic CNT kinetic energy (as estimated by the inverse density of states). This is a prerequisite for a tunnel barrier, although the ratio of height to kinetic energy suggests a decay length not much smaller than the barrier width itself, and therefore a relatively high transmission probability. Note that our first-principles calculations also help explain the large MR because they indicate (Fig. 3b) that the LSMO surface is highly spin polarized despite a pronounced interfacial state ~0.2 eV below the Fermi level.

Our main result is the observation of a large device MR (61% in Fig. 4) that arises due to sharp and irreversible switching of the LSMO electrode magnetizations between P and AP configurations. Three other working devices showed 54%, 72% and 53% (Supplementary Information). These four MR values are much higher than the ~10% values observed with CNTs between other ferromagnetic contacts[12-15]. We now discuss why the use of a CNT in place of a standard semiconductor permits the large MR.

The MR of a structure composed of a conduction channel connected to a ferromagnetic source and drain through spin dependent interface resistances (e.g. a tunnel junction) can be expressed[21,22] as:

$$MR = \frac{\Delta R}{R_P} \equiv \frac{R_{AP} - R_P}{R_P} \equiv \frac{\gamma^2/(1-\gamma^2)}{1+\tau_n/\tau_{sf}}, \quad (1)$$

where $\gamma$ is the electrode spin polarization, or more formally the interfacial spin asymmetry coefficient on which the spin dependent interface resistance $r_{\uparrow(\downarrow)} = 2(1\mp\gamma)r_b^*$ depends, $r_b^*$ is the mean value of the spin independent interface resistance, $\tau_{sf}$ is the spin lifetime and $\tau_n$ is the dwell time of the electrons in the channel:

$$\tau_n = 2L/(v_N \bar{t}_r). \quad (2)$$

$L$ is the length of the channel, $v_N$ is the mean electron velocity in the channel (here, $v_F$ for the CNT), and $\bar{t}_r$ is the mean transmission coefficient at each interface (related to the interfacial resistance by a classical Landauer equation – see later). Equations (1) and (2) hold for ballistic transmission from source to drain, and also for diffusive transport when the interface resistance is sufficiently large[21,22], as we have here.

The key point in Equation (1) is that the MR is controlled by two factors: trivially $\gamma$, but critically also the ratio $\tau_n/\tau_{sf}$. If this ratio is large, the MR will tend to zero whatever the value of $\gamma$. From Equation (2) we can express this ratio as:

$$\frac{\tau_n}{\tau_{sf}} = \frac{2L}{v_N \bar{t}_r \tau_{sf}}. \qquad (3)$$

Purely metallic structures like magnetic multilayers have the advantage of a large $v_N$ and a large $\bar{t}_r \sim 1$, but $\tau_{sf}$ is very short so that a large $\Delta R/R$ can be obtained only when $L$ is short, e.g. in current-perpendicular-to-the-plane giant magnetoresistance. The long $L$ in a lateral structure forces $\Delta R/R$ to become small, e.g.[1] ~5%. When the interfaces are tunnel junctions, in vertical as well as lateral structures, the concomitant reduction of $\bar{t}_r$ leads to an even smaller $\Delta R/R$, e.g.[23] ~$10^{-4}$.

With semiconductors, the possible advantage is a long[1] $\tau_{sf}$, but the carrier velocity is small. For example, in n-type GaAs ($10^{17}$ cm$^{-3}$), for which long conduction band spin lifetimes of several nanoseconds have been found at low temperatures, the mean velocity along a channel axis can be estimated to be $3 \times 10^4$ m/s, c.f. $10^6$ m/s in metals or CNTs. Also with semiconductors, $\bar{t}_r$ is small given the tunnel barrier required for spin injection from a metal[24-26]. Therefore, to our knowledge, a significant electrical signal, MR $\approx$ 40%, has only be observed with the very small $L$ ($\approx$ 5-10 nm) of vertical structures[27] that are unsuitable for gating. As noted above, the MR does not exceed 1% in lateral structures[2].

The advantage of CNTs is that they combine the long $\tau_{sf}$ of semiconductors with the large[8] $v_F$ of metals. Therefore the MR is large here despite our long $L = 2$ μm, and our small $\bar{t}_r$ (calculated later) associated with the interfacial tunnel barriers. We now proceed to a quantitative discussion of our results. From Equation (1), $\gamma$ and $\tau_n/\tau_{sf}$ cannot be extracted from the MR alone (61% at 5K and 25 mV), but we necessarily have $\gamma \geq 0.62$ since the denominator of Equation (1) cannot be smaller than unity. It is possible that $\gamma=1$ for half-metallic LSMO, but interfacial imperfections always lead to smaller values. The maximum value that has been observed in epitaxial magnetic tunnel junctions[4] with LSMO is 0.95 and other experiments have found smaller values. Here we propose a tentative scenario assuming a reasonable value of $\gamma=0.8$. Using this figure in Equation (1), the experimental MR of 61% gives $\tau_n/\tau_{sf} \cong 2$.

Separately, we find $\tau_n \cong 60$ ns from Equation (2), using $L = 2$ μm, $v_F = 0.8 \times 10^6 m/s$ [8] and a value of $\bar{t}_r$ estimated from the interface resistance $r_b^*$ using the Landauer equation:

$$r_b^* = \frac{h}{4e^2 \bar{t}_r}. \qquad (4)$$

The assumption of two spin-degenerate conduction channels in Equation (4) is reasonable[8] even for a multiwall CNT. As the interfaces dominate device resistance $R$, we have taken $r_b^* = R/2 \cong 75$ MΩ. This leads to $\bar{t}_r \cong 0.9 \times 10^{-4}$, $\tau_n \cong 60$ ns, and finally, from our previous estimate of $\tau_n/\tau_{sf} \cong 2$, to $\tau_{sf} \cong 30$ ns.

Our findings turn out to be very reasonable given the very weak spin-orbit coupling of carbon, and should therefore be relevant to a wide family of carbon-based molecules. Regarding the spin diffusion length, if we assume that the transport in the CNT is diffusive with a mean free path[8] λ~100 nm, our value of $\tau_{sf}$ corresponds to $l_{sf} = \sqrt{v_F \tau_{sf} \lambda} \cong 50 \mu m$. The same calculation with the best value of γ~0.95 observed in epitaxial LSMO tunnel junctions[4] would reduce $\tau_n$ by a factor of 7 and shorten $l_{sf}$ by a factor of 2.7.

Device MR falls with increasing temperature (Fig. 5), but the field dependence is qualitatively unchanged. Our MR is halved at 40 K, and persists to 120 K which although well below room temperature is a significant improvement on previous molecular spintronics devices[10-15]. This loss of performance well below the 365 K Curie temperature of bulk LSMO is likely associated with the well known thermal suppression of spin polarization[3]. A similar fall off in performance seen in LSMO tunnel junctions[29] is attributed to a reduced interfacial Curie temperature arising from charge transfer or loss of bulk symmetry. Replacing LSMO with a high Curie temperature metal such as Co could solve this problem, but the poor results obtained previously[12-15] suggest the need for appropriate interfacial tunnel barriers, e.g. thin insulating layers, in order to be able to work at high voltage without unduly increasing the current.

The decrease of MR with increasing bias at 5 K (Fig. 5) is also reminiscent of LSMO tunnel junctions. Above the classical zero bias anomaly, not resolved here, a plateau out to ~110 mV is seen, and then the decrease is steep. However, we cannot rule out the possibility that this bias dependence is influenced by a mechanism related to the energy band structure of the CNT. The persistence of the MR plateau out to ~110 mV permits the associated output signal, i.e. the voltage difference between the P and AP configurations for the same current, to increase from 15 mV at a bias of 25 mV up to 65 mV at a bias of 110 mV. This figure of 65 mV falls in a suitable voltage range for applications.

We have thus demonstrated that spin injection using a CNT with LSMO electrodes can be used to transform spin information into a significant output voltage. As explained earlier, the advantage of a CNT is that it combines the long $\tau_{sf}$ of some semiconductors, with a very high $v_F$ such that $\tau_n$ is low (Equations 1&2). Moreover, the tunnel barriers at our LSMO-CNT interfaces are low enough to keep $\tau_n \sim \tau_{sf}$, but large enough to limit the current at our high bias voltages that, in order to avoid albeit very interesting complexities, lie well above the energy splitting induced by Coulomb blockade and quantization effects. Our first-principles calculations support the existence and intrinsic nature of the observed barrier, and also the persistence of the high bulk spin polarization at the interface which is a prerequisite for the large MR observed (Equation 1).

Our work forms part of the nascent molecular spintronics paradigm in which it is possible to manipulate spin polarized electrons in novel environments. However, the weak spin-orbit coupling in carbon precludes the electrically driven magnetic reversal of spins in a CNT-based Datta and Das[5] spin transistor. Instead, the spin precession induced by the stray magnetic field of a ferromagnetic gate, i.e. the Hänle effect[1], could be used to flip spins in a CNT. Given that the precession angle induced by a transverse field $B$ during time $t$ is $2\mu_B B t / \hbar$, our $\tau_n$~60 ns suggests that the application of a modest 10 mT to a small fraction of the length of a CNT (a few tenths of microns) would be sufficient to reverse the spin polarization between injection and detection. In future, one might seek non-magnetic channels with intermediate levels of spin-orbit coupling in order to permit spin manipulation by the electric field of a gate without unduly reducing the spin lifetime and the output signals.

**METHODS**
**Experimental**

Epitaxial LSMO thin films were grown on closely lattice matched orthorhombic NdGaO$_3$ (001) substrates by pulsed laser deposition with a KrF excimer laser (248 nm, 1 Hz, 2.5 Jcm$^{-2}$, 775$^o$C, 15 Pa O$_2$, target-substrate distance = 8 cm). The films display step-terrace growth, and possess in-plane uniaxial magnetocrystalline anisotropy in the orthorhombic [100] direction. Below 360 K the films are ferromagnetic (3.6 $\mu_B$/Mn at 10 K), and on cooling the resistivity decreases to ~60 $\mu\Omega$.cm at 10 K. Using conventional photolithography, electrode tracks (widths 1-4 µm, separation 1.5 µm) were defined perpendicular to [100], so that their magnetizations could be switched independently by an external magnetic field. Multiwall CNTs of diameter ~20 nm grown by arc-discharge (Iljin Nanotech Co. Ltd., Korea) were subsequently dispersed from a 1,2-dicloroethane solution. A SEM was used to confirm the presence of a single nanotube running between adjacent electrically connected electrodes. Electrical measurements of interest were made using a Keithley source meter in constant voltage mode.

**Theoretical**

First-principles electronic-structure calculations were performed within the density-functional-theory (DFT) framework[20] in the spin-polarised generalized-gradient approximation, using the SIESTA method[29]. Further details on the performance of the method for LSMO can be found elsewhere[30]. The MnO$_2$-terminated (001) surface of LSMO was described by a 23-layer slab of LSMO, in which one third of the La atoms where replaced[30] by Sr. A (6,6) single-wall CNT was put onto the LSMO surface in a commensurate arrangement in which 3 unit cells of the CNT were laid along the (100) direction on a 4x2 lateral supercell of LSMO. The mismatch strain is 5 %. The atomic positions of the CNT on the previously relaxed surface were obtained by minimizing the mutual DFT forces. Even though experiments were performed on multiwall nanotubes which are arguably better described in the graphitic limit, we have nevertheless considered a nanotube, since the dimensionality greatly affects the contact resistance, and the qualitative picture emerging from the calculations should remain.

**Acknowledgements.** We thank G. A. J. Amaratunga, H. Bouchiat, L. Brey, M. R. Buitelaar, M. J. Calderón, S. N. Cha, M. Chhowalla, A. Cottet, H. Jaffrès, D.-J. Kang, T. Kontos, P. Seneor and N. A. Spaldin. This work was funded by the UK EPSRC, NERC, BNFL, The Royal Society, Spanish MEC (JMP), Donostia International Physics Center (EA) and the EU.

**Corresponding author:** N.D.M. (ndm12@cam.ac.uk)

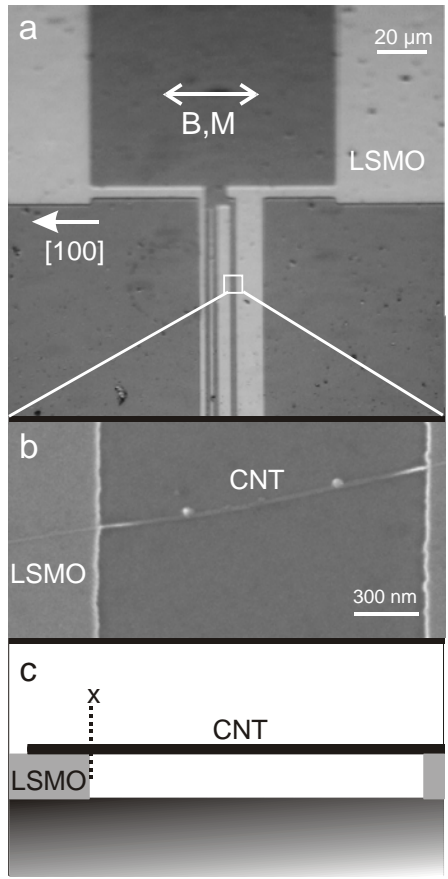

**Figure 1 | LSMO-CNT-LSMO device. a,** Optical micrograph of four variable width LSMO electrodes, and two of the four associated contact pads. In electrically conducting devices, two adjacent electrodes were connected by an overlying CNT, in regions such as the one in the white square. Magnetic fields *B* were applied along the orthorhombic [100] direction in which the magnetization *M* is expected to lie due to uniaxial magnetocrystalline anisotropy. **b,** SEM image of a CNT running between LSMO electrodes. **c,** schematic side view of **b** with the plane through the CNT at the edge of the LSMO electrode denoted ×.

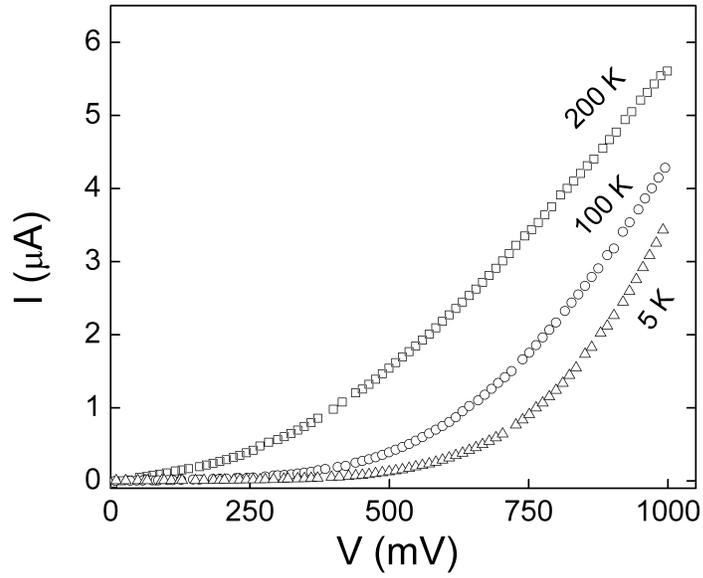

**Figure 2 | Zero-field non-linear *I-V* characteristics for a typical working LSMO-CNT-LSMO device.** The data correspond to the low resistance state associated with the parallel magnetic electrode configuration, and are symmetric about *V*=0 as expected from device symmetry.

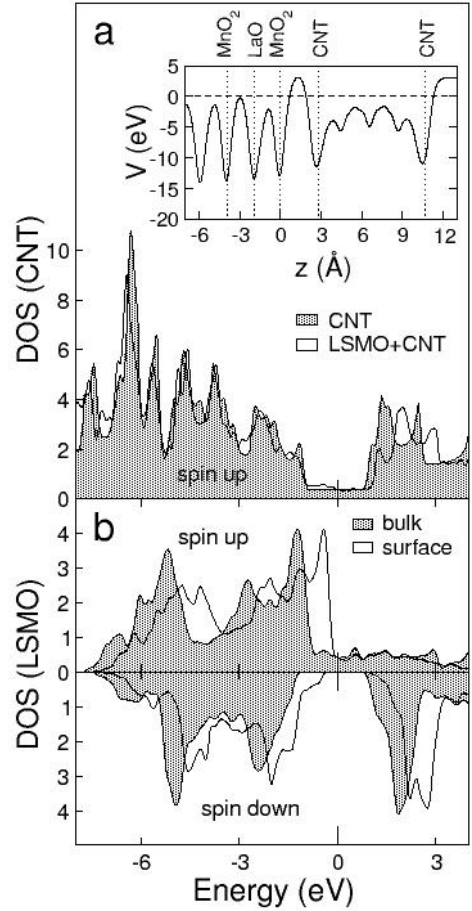

**Figure 3 | First-principles calculations of device interfaces.** Projected density of states (DOS) on **a,** the basis functions of an isolated CNT (shaded), and a CNT lying on LSMO (unshaded). **b,** the projected DOS onto the first $MnO_2$+(La,Sr)O layer of the LSMO slab (unshaded) and onto bulk LSMO (shaded). Fermi levels aligned at zero energy, and only up spins shown in **a** since up-down differences in the CNT DOS are barely visible at this scale (there is a net spin polarisation of +0.01 electrons/Å). Inset, the Kohn-Sham potential seen by electrons in the vicinity of the LSMO/CNT interface. It has been integrated for each value of *z* (normal to the LSMO surface) in the rectangle defined by the projection of the CNT onto the *x-y* plane. The origin of potential has been chosen at the Fermi level (horizontal dashed line). Vertical dotted lines indicate the nuclear positions of the atomic layers of LSMO, and the limits of the CNT.

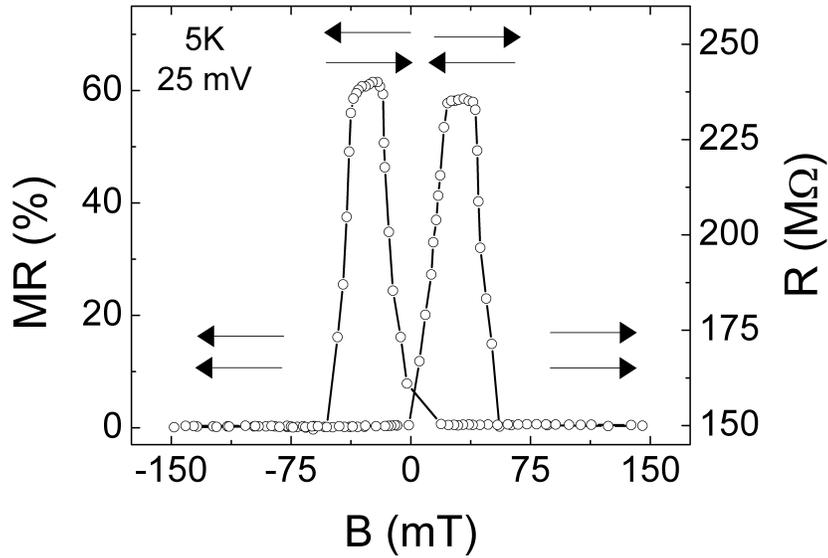

**Figure 4 | LSMO-CNT-LSMO device MR.** Data recorded at 5 K with a bias voltage of 25 mV show two distinct states of resistance *R*, as the magnetic configuration of the two LSMO electrodes is switched by an applied magnetic field *B*. The arrows indicate the relative magnetic orientation of the electrodes, which possess different switching fields because of their different widths. The data points and interconnecting lines were generated by averaging over 25 cycles. In Supplementary Information, we show similar MR data for three other working devices. One of these three devices was fabricated with silica between the manganite electrodes to prevent the possibility of the CNT sagging. For another of these three devices, we show data from a single field sweep. $MR(\%)=100\times[R(B)-R(0)]/R(B)$.

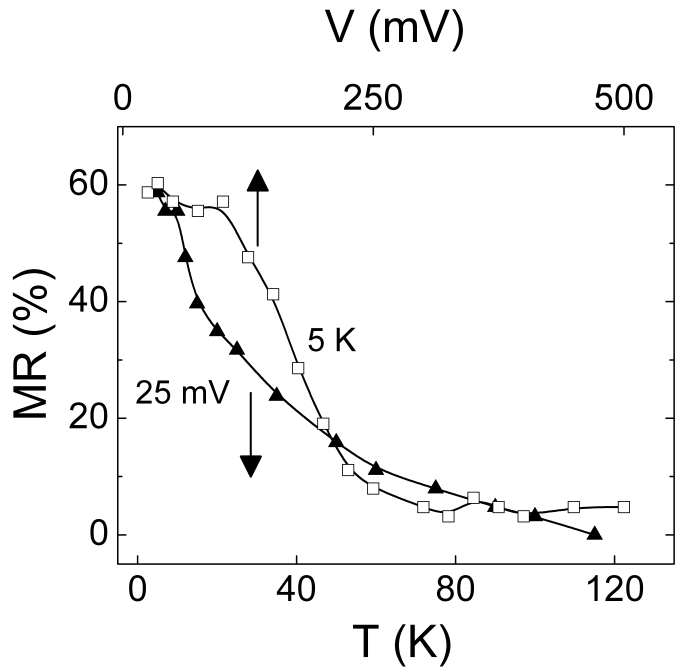

**Figure 5 | Temperature and bias dependence of peak MR.** The magnitude of the two-state switching seen in Fig. 4 is plotted as a function of bias V at low temperature (□), and as a function of temperature $T$ at 25 mV (▲). $MR(\%)=100\times[R_{AP}-R_P]/R_P$.

# Supplementary information

# Transformation of spin information into large electrical signals via carbon nanotubes

Luis E. Hueso, José M. Pruneda, Valeria Ferrari, Gavin Burnell, José P. Valdés-Herrera, Benjamin D. Simons, Peter B. Littlewood, Emilio Artacho, Albert Fert & Neil D. Mathur

Similar and reproducible MR effects were seen in 4 devices, in which we suggest that magnetic impurities in the CNT are absent. These 4 devices formed a subset of the 12 devices that showed similar and reproducible current-voltage (*I-V*) characteristics, and these 12 devices formed a subset of the 60 devices that showed some form of electrical conductivity. Device 1 was used for Figures 4 and 5 of our manuscript, and here we reproduce that data alongside corresponding data for Devices 2-4. For Device 3, we show *MR(B)* for a single field sweep whereas an average over 25 cycles was used for the other devices. Device 2 was fabricated with silica between the manganite electrodes to prevent the possibility of the CNT sagging, but no significant changes were seen.

*P.T.O.*

## Device 1. MR (5 K, 25 mV) = 61%

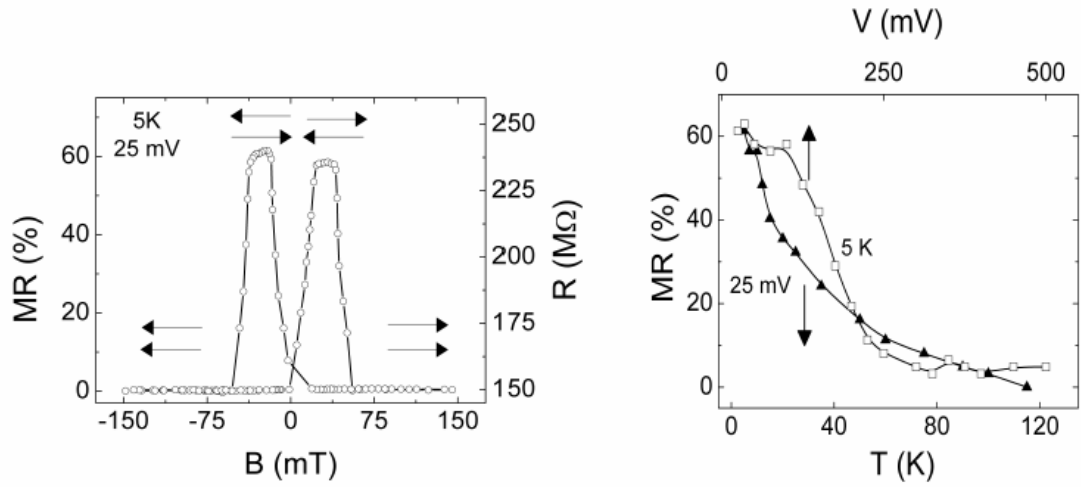

## Device 2. MR (5 K, 25 mV) = 54%.
### SiO$_2$ between LSMO electrodes

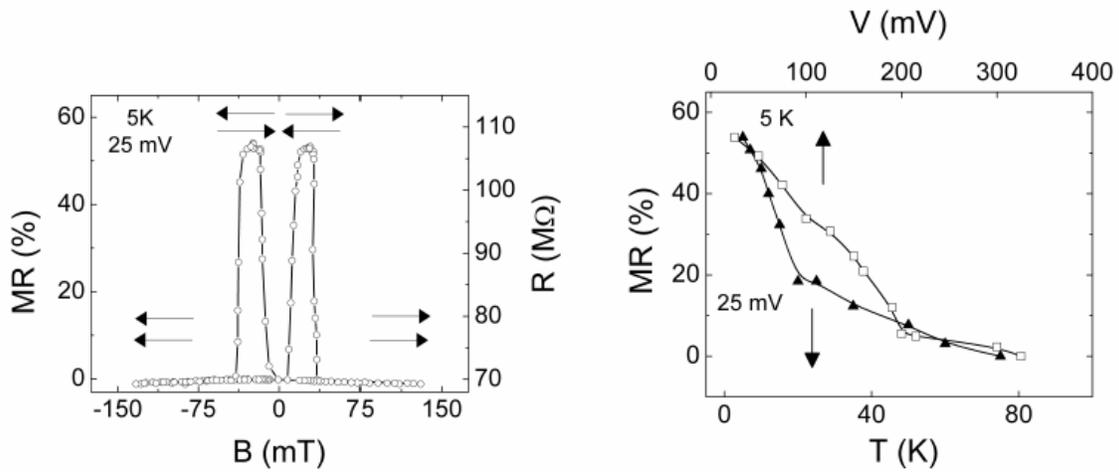

*P.T.O.*

## Device 3. MR (5 K, 25 mV) = 72%

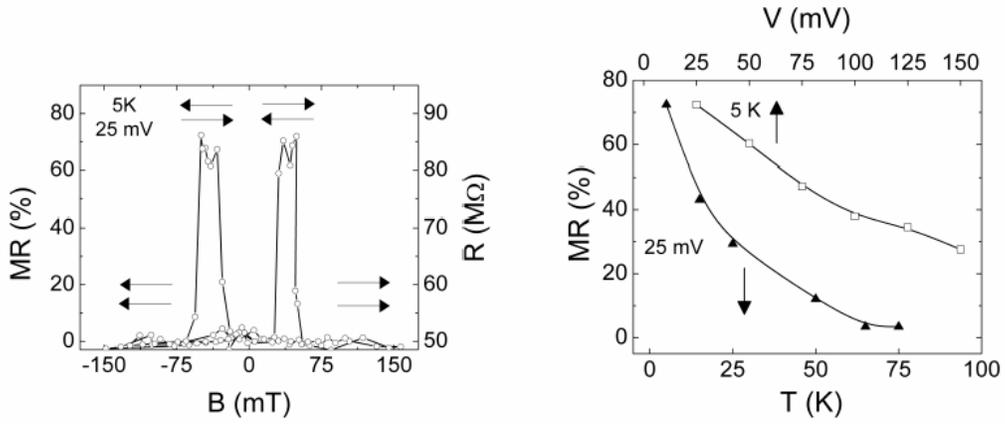

## Device 4. MR (5 K, 25 mV) = 53%

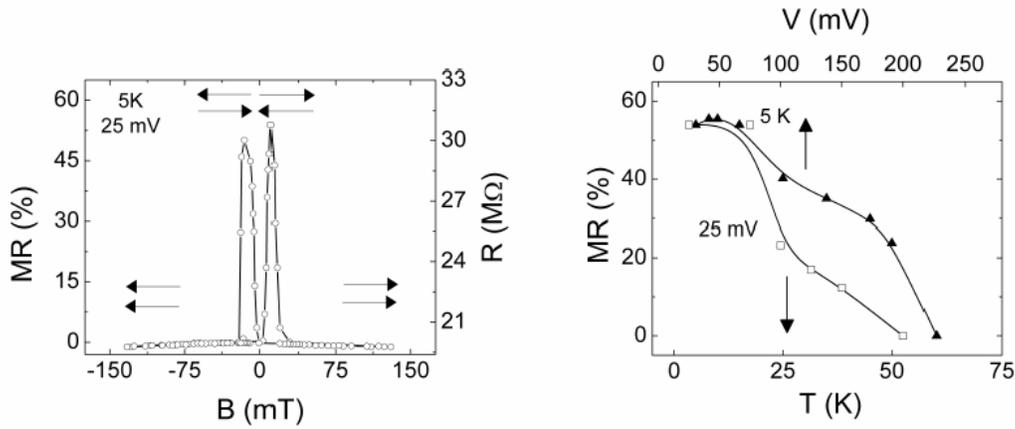

*P.T.O.*